\documentclass{article}
\usepackage{makeidx}  
\usepackage{amsmath}  
\usepackage{amsfonts}
\usepackage{authblk}
\usepackage[a4paper, total={6in, 8in}]{geometry}
\usepackage{natbib}

\DeclareMathOperator*{\argmax}{arg\,max}

\begin{document}

\title{Estimating probabilistic context-free grammars for proteins using contact map constraints}

\author{Witold Dyrka\thanks{Corresponding author: witold.dyrka@pwr.edu.pl}}
\affil{Politechnika Wroc\l awska, Wydzia\l\ Podstawowych Problem\'ow Techniki, Katedra In\.zynierii Biomedycznej, Poland}
\author{Fran\c{c}ois Coste}
\author{Juliette Talibart}
\affil{Irisa / Inria Rennes - Bretagne Atlantique, France}

\date{}

\maketitle

\begin{abstract}
Learning language of protein sequences, which captures non-local interactions between amino acids close in the spatial structure, is a long-standing bioinformatics challenge, which requires at least context-free grammars. However, complex character of protein interactions impedes unsupervised learning of context-free grammars. Using structural information to constrain the syntactic trees proved effective in learning probabilistic natural and RNA languages. In this work, we establish a framework for learning probabilistic context-free grammars for protein sequences from syntactic trees partially constrained using amino acid contacts obtained from wet experiments or computational predictions, whose reliability has substantially increased recently. Within the framework, we implement the maximum-likelihood and contrastive estimators of parameters for simple yet practical grammars. Tested on samples of protein motifs, grammars developed within the framework showed improved precision in recognition and higher fidelity to protein structures. The framework is applicable to other biomolecular languages and beyond wherever knowledge of non-local dependencies is available.
\emph{Keywords}: probabilistic context-free grammar, syntactic tree, structural constraints, protein sequence, protein contact map, maximum-likelihood estimator, contrastive estimation
\end{abstract}

\section{Introduction}

\subsection{Grammatical modeling of proteins}
The essential biopolymers of life, nucleic acids and proteins, share the basic characteristic of the languages: infinite number of sequences can be expressed with a finite number of monomers. In the case of proteins, merely 20 amino acids species (letters) build millions of sequences (words or sentences) folded in thousands of different spatial structures playing various functions in living organisms (semantics). Physically, the protein sequence is a chain of amino acids linked by peptide bonds. The physico-chemical properties of amino acids and their interactions across different parts of the sequence defines its spatial structure, which in turn determines biological function to great extent. Similarly to words of the natural language, protein sequences may be ambiguous (the same amino acid sequence folds into different structures depending on the environment), and often include non-local dependencies and recursive structures \citep{searls2013}. 

Not surprisingly the concept of \textit{protein language} dates back to at least 1960s \citep{Pawlak1965}, and since early applied works in 1980s \citep{bre84, jim84} formal grammatical models have gradually gained importance in bioinformatics \citep{se02, searls2013, cos16}. Most notably, Hidden Markov Models (HMM), which are weakly equivalent to probabilistic regular grammars, became the main tool of protein sequence analysis. Profile HMM are commonly used for defining protein families \citep{son98, Finn2015} and for searching similar sequences \citep{edd98, Eddy2011, soe05a, rem12}; and more expressive HMM are developed \citep{cos06,bre12}. Yet, their explanatory power is limited since, as regular level models, they cannot capture non-local interactions, which occur between amino acids distant in sequence but close in the spatial structure of the protein. Many of these interactions have a character of nested, branched and crossing dependencies, which in terms of grammatical modelling requires context-free (CF) and context-sensitive (CS) level of expressiveness \citep{searls2013}. However, grammatical models beyond regular levels have been rather scarcely applied to protein analysis (a comprehensive list of references can be found in \citep{Dyrka2013}. This is in contrast to RNA modeling, where CF grammatical frameworks are well developed and power some of the most successful tools \citep{sa93, Eddy1994, kn99, Sukosd2012}. 

One difficulty with modeling proteins is that interactions between amino acids are often less specific and more \textit{collective} in comparison to RNA. Moreover, the larger alphabet made of 20 amino acid species instead of just 4 bases in nucleic acids, combined with high computational complexity of CF and CS grammars, impedes inference, which may lead to solutions which does not outperform significantly HMMs \citep{dy09, Dyrka2013}. Yet, some studies hinted that CF level of expressiveness brought an added value in protein modeling when grammars fully benefiting from CF nesting and branching rules were compared in the same framework to grammars effectively limited to linear (regular) rules \citep{dy07, Dyrka2013}. Good preliminary results were also obtained on learning sub-classes of CF grammars to model protein families, showing the interest of taking into account long distance correlations in comparison to regular models \citep{cos12,cos14}.

An important advantage of CF and CS grammars is that parse trees they produce are human readable descriptors. In RNA modeling, the shape of parse trees can be used for secondary structure prediction \citep{Dowell2004}. In protein modeling, it was suggested that the shape of parse trees corresponds to protein spatial structure \citep{dy09}, and that they can also convey biologically relevant information \citep{sci11,Dyrka2013}. 

\subsection{Grammar estimation with structural constraints}
In this piece of research the focus is on learning probabilistic context-free grammars (PCFG) \citep{Booth1969}. Learning PCFG consists in estimating the unfixed parameters of the grammar with the aim of concentrating probability mass from the entire space of possible sequences and their syntactic trees to the target population, typically represented by a sample. The problem is often confined to assigning probabilities to fixed production rules of a generic underlying non-probabilistic CFG \citep{Lari1990}. Typically the goal is to estimate the parameters to get a grammar maximizing the likelihood of the (positive) sample, while, depending on the target application, other approaches also exists. For example, the contrastive estimation aims at obtaining grammars discriminating target population from its neighbourhood \citep{Smith2005}. 

The training sample can be made of a set of sequences or a set of syntactic trees. In the former case, all derivations for each sentence are considered valid. Given the underlying non-probabilistic CFG, probabilities of rules can be estimated from sentences in the classical Expectation Maximization framework (e.g. the Inside-Outside algorithm \citep{ba79, Lari1990}), however, the approach is not guaranteed to find the globally optimal solution \citep{Carroll1992}. Heuristic methods applied for learning PCFG from positive sequences include also iterative biclustering of bigrams \citep{Tu2008}, and genetic algorithms using a learnable set of rules \citep{Kammeyer1996, Keller1998, Keller2005} or a fixed covering set of rules \citep{Tariman2004, dy09}.

Much more information about the language is conveyed in the syntactic trees. If available, a set of trees (a treebank) can be directly used to learn a PCFG \citep{Charniak1996}. Usability of structural information is highlighted with the result showing that a large class of non-probabilistic CFG can be learnt using unlabeled syntactic trees (called also \textit{skeletons}) of the training samples \citep{Sakakibara1992}. Algorithms for learning probabilistic CF languages, which exploits structural information in syntactic trees, have been proposed \citep{sa93, Eddy1994, Carrasco2001, Cohen2014}. An interesting middle way between plain sequences and syntactic trees are partially bracketed sequences, which constrain the shape of the syntactic trees (the skeletons) but not node labels. The approach was demonstrated to be highly effective in learning natural languages \citep{Pereira1992}. It was also applied to integrating uncertain information on pairing of nucleotides of RNA \citep{Knudsen2005}. In this approach the modified bottom-up parser penalizes probability on derivations inconsistent with available information on nucleotide pairing in such way that the amount of the penalty is adjusted according to confidence of the structural information. 

\subsection{Protein contact constraints}
\label{sec:intro-protein-contacts}
To our knowledge constrained sets of syntactic trees have never been applied for estimating PCFG for proteins. In this research we propose to use spatial contacts between amino acids distant in sequence as a source of constraints. Indeed, an interaction between amino acids, which forms a dependency, usually requires a contact between them, defined as spatial proximity. Until recently, extensive contact maps were only available for proteins with experimentally solved structures, while individual interactions could be determined through mutation-based wet experiments. 

Currently, reasonably reliable contact maps can also be obtained computationally from large collective alignments of evolutionary related sequences. The rationale for the contact prediction is that if amino acids at a pair of positions in the alignment interact then a mutation at one position of the pair often requires a compensatory mutation at the other position in order to maintain the interaction intact. Since only proteins maintaining interactions vital for function successfully endured the natural selection, an observable correlation in amino acid variability at a pair of positions is expected to indicate interaction. However, standard correlations are transitive and therefore cannot be immediately used as interaction predictors. The break-through was achieved recently by Direct Coupling Analysis (DCA)\citep{wei09}, which disentangles direct from indirect correlations by inferring a model on the alignment which can give information on the interaction strengths of the pairs. There are different DCA methods based on how the model, which is usually a type of the Markov Random Field, is obtained \citep{mor11, jon12, ekeberg2013, kamisetty2013, Seemayer2014, baldassi2014}. The state-of-the-art DCA-based meta-algorithms achieve mean precision in the range 42-74\% for top $L$ predicted contacts and 69-98\% for top $L/10$ predicted contacts, where $L$ is the protein length \citep{Wang2017}. Precision is usually lower for shorter sequences and especially for smaller alignments, however a few top hits may still provide relevant information \citep{Daskalov2015}.

\subsection{Structure of the document}
The rest of the document is organized as follows. Section~\ref{sec:methods} introduces the main contribution of this work. First, a novel PCFG-CM framework for learning PCFG with the Contact-Map constraints is established, for which the maximum-likelihood and contrastive estimators are defined (section~\ref{sec:general-model}). Second, a special instance of the problem, both simple and practical, is considered: suitable forms of the grammar and the contact constraints are defined, a variant of the bottom-up chart parser is proposed, and effective calculations of one of the contrastive estimators for the proposed form of grammar are given (section~\ref{sec:simple-instance}). The setup of experimental evaluation of the PCFG-CM approach for the special instance is described in section~\ref{sec:evaluation}. Section~\ref{sec:results} presents sample data and results of evaluation. Eventually, section~\ref{sec:discussion} concludes the document with discussion of the results, limitations and perspectives for future work.

\section{Methods}
\label{sec:methods}

\subsection{General model}
\label{sec:general-model}

\subsubsection{Basic notations}
\label{sec:basic-notations}
Let $\Sigma$ be a non-empty finite set of atomic symbol (representing for instance amino acid species). The set of all finite strings over this alphabet is denoted by $\Sigma^*$. Let $|x|$ denote the length of a string $x$. The set of all strings of length $n$ is denotes by $\Sigma^n = \{x \in \Sigma^* \colon |x|=n\}$. Let $x = x_1 \ldots x_n$ be a sequence in $\Sigma^n$. 

\paragraph{Unlabeled syntactic tree}
An unlabeled syntactic tree (UST) $u$ for $x$ is an ordered rooted tree such that the leaf nodes are labeled by $x$, which is denoted as $yield(u) = x$, and the non-leaf nodes are unlabeled. Let $\mathcal{U}_*$ denotes the set of all USTs that yield a sequence in $\Sigma^*$, let $\mathcal{U}_n = \{u\in\mathcal{U}_*: yield(u) \in \Sigma^n\}$, where $n$ is a positive integer, and let $\mathcal{U}_x = \{u\in\mathcal{U}_*: yield(u)=x\in\Sigma^*\}$. Note that $\forall(x,w\in\Sigma^*,\ x \neq w)\ \mathcal{U}_x \cap \mathcal{U}_w = \emptyset$ and $\mathcal{U}_* = \cup_{x \in \Sigma^*} \mathcal{U}_x$. Moreover, let $U$ denotes an arbitrary subset of $\mathcal{U}_*$.

\paragraph{Context-free grammar}
A context-free grammar (CFG) is a quadruple $G=\langle \Sigma, V, v_0, R \rangle$, where $\Sigma$ is defined as above, $V$ is a finite set of non-terminal symbols (also called variables) disjoint from $\Sigma$, $v_0 \in V$ is a special start symbol, and $R$ is a finite set of rules rewriting from variables into strings of variables and/or terminals $R = \{r_i \colon V \rightarrow (\Sigma \cup V)^*\}$. Let $\alpha=\alpha_1\ldots\alpha_k$ be a sequence of symbols in $(\Sigma \cup V)^k$ for some natural $k$. A (left-most) derivation for $G$ is a string of rules $r=r_1\ldots r_l \in R^l$, which defines an ordered parse tree $y$ starting from the root node labeled by $v_0$. In each step, by applying a rule $r_i: v_j \rightarrow \alpha_1 \ldots \alpha_k$, tree $y$ is extended by adding edges from the already existing left-most node labeled $v_j$ to newly added nodes labeled $\alpha_1$ to $\alpha_k$. Therefore there is a one-to-one correspondence between derivation $r$ and parse tree $y$. Derivation $r$ is complete if all leaf nodes of the corresponding (complete) parse tree $y$ are labeled by symbols in $\Sigma$. Sets $\mathcal{Y}_*$, $\mathcal{Y}_n$ and $\mathcal{Y}_x$ are defined as for the USTs. For a given parse tree $y$, $u(y)$ denotes the unlabeled syntactic tree obtained by removing the non-leaf labels on $y$. Given a UST $u$, let $\mathcal{Y}_G(u)$ be the set of all parse trees for grammar $G$ such that $u(y)=u$. For a set of USTs $U$, $\mathcal{Y}_G(U) = \cup_{u \in U} \mathcal{Y}_G(u)$. Note that $\forall(u, v \in U,\ u \neq v)\ \mathcal{Y}_G(u) \cap \mathcal{Y}_G(v) = \emptyset$. 

\paragraph{Probabilistic context-free grammar}
A probabilistic context-free grammar (PCFG) is a quintuple $\mathcal{G}=\langle \Sigma, V, v_0, R, \theta \rangle$, where $\theta$ is a finite set of probabilities of rules: $\theta = \{\theta_i=\theta(r_i): R \rightarrow [0,1]\}$, setting for each rule $v_k\rightarrow \alpha$ its probability to be chosen to rewrite $v_k$ with respect to other rules rewriting $v_k$ (such that $\forall(v_k \in V)\ \sum_{v_k \rightarrow \alpha} \theta(v_k \rightarrow \alpha) = 1$). Let PCFG $\mathcal{G}$ that enhances the underlying non-probabilistic CFG $G = \langle \Sigma, V, v_0, R \rangle$ is denoted by $\mathcal{G}=\langle G, \theta \rangle$.
The probability of parse tree $y$ using the probability measure induced by $\mathcal{G}$ is given by the probability of the corresponding derivation $r=r_1 \ldots r_l$:
$$prob(y \mid \mathcal{G}) = prob(r \mid \mathcal{G}) = \prod_{i=1}^l \theta(r_i).$$
$\mathcal{G}$ is said to be \textit{consistent} when it defines probability distribution over $\mathcal{Y}_*$:
$$prob(\mathcal{Y}_* \mid \mathcal{G}) = \sum_{y \in \mathcal{Y}_*} prob(y \mid \mathcal{G})  = 1.$$
The probability of sequence $x \in \Sigma^*$ given $\mathcal{G}$ is:
$$ prob(x \mid \mathcal{G}) = prob(\mathcal{Y}_x \mid \mathcal{G}) = \sum_{y \in \mathcal{Y}_x} prob(y \mid \mathcal{G}),$$
and the probability of UST $u \in \mathcal{U}_x$ given $\mathcal{G}$ is:
$$prob(u \mid \mathcal{G}) = prob(\mathcal{Y}_G(u) \mid \mathcal{G}) = \sum_{y \in \mathcal{Y}_G(u)} prob(y \mid \mathcal{G}).$$
Since $\mathcal{Y}_x$ and $\mathcal{Y}_G(u)$ define each a partition of $\mathcal{Y}_*$ for $x\in \Sigma^*$ and for $u\in \mathcal{U}_*$, a consistent grammar $\mathcal{G}$ defines also a probability distribution over $\Sigma^*$ and  $\mathcal{U}_*$.

\subsubsection{Contact constraints} 
\label{sec:contact-constraints}
Most proteins sequences fold into complex spatial structures. Two amino acids at positions $i$ and $j$ in the sequence $x$ are said to be in contact if distance between their coordinates in spatial structure $d(i,j)$ is below a given threshold $\tau$. A full contact map for a protein of length $n$ is a binary symmetric matrix $\mathsf{m}^\mathrm{full} = (m_{i,j})_{n \times n}$ such that $m_{i,j}=[d(i,j)<\tau]$, where $[x]$ is the Iverson bracket. Usually only a subset of the contacts is considered (cf section~\ref{sec:intro-protein-contacts}). A (partial) contact map for a protein of length $n$ is a binary symmetric matrix $\mathsf{m} = (m_{i,j})_{n \times n}$ such that $m_{i,j}=1 \implies d(i,j)<\tau$. Let $d_u(i,j)$ is the length of the path from $i$-th to $j$-th leaf in UST $u$ for $x$. Given a threshold $\delta$, UST $u$ is said to be consistent with a contact map $\mathsf{m}$ of length $n$ if $m_{i,j}=1 \implies d_u(i,j) < \delta$.

For a contact map $\mathsf{m}$ of length $n$, let $\mathcal{U}^\mathsf{m}_n$ denotes the subset of $\mathcal{U}_n$ consistent with $\mathsf{m}$, and $\mathcal{U}^\mathsf{m}_x$ denotes the subset of $\mathcal{U}_x$ consistent with $\mathsf{m}$. Note that $\mathcal{U}^\mathsf{m}_x = \mathcal{U}^\mathsf{m}_n \cap \mathcal{U}_x$. Analogous notations apply to parse trees.

\subsubsection{Estimation}
\label{sec:estimation}
Learning grammar $\mathcal{G} = \langle \Sigma, V, v_0, R, \theta \rangle$ can be seen as estimating the unfixed parameters of $\mathcal{G}$ with the aim of concentrating probability mass from the entire space of unlabeled syntactic trees $\mathcal{U}_*$ to the set of unlabeled syntactic trees for the target population $\mathfrak{U}_\mathrm{target}$. In practice, only a sample of the target population can be used for learning, hence estimation is performed on $\mathfrak{U}_\mathrm{sample} \subseteq \mathfrak{U}_\mathrm{target}$. Note that even in the most general case the set of terminal symbols $\Sigma$ is implicitly determined by the sample; moreover the start symbol $v_0$ is typically also fixed. A common special case confines learning grammar $\mathcal{G}$ to estimating $\theta$ for a fixed quadruple of non-probabilistic parameters $\langle \Sigma, V, v_0, R \rangle$ (which fully determine the non-probabilistic grammar $G$ underlying $\mathcal{G}$). Given inferred grammar $\mathcal{G}_*$ and a query set of unlabeled syntactic trees $\mathcal{U}_\mathrm{query}$, probability $prob(\mathcal{U}_\mathrm{query} \mid \mathcal{G}_*)$ is an estimator of the likelihood that $\mathcal{U}_\mathrm{query}$ belongs to population $\mathfrak{U}_\mathrm{target}$. 

\paragraph{Maximum likelihood grammar}
Let $X$ be a sample set of sequences in $\Sigma^*$, and let $\mathsf{M}$ be a set of corresponding contact matrices. The sample set $\mathcal{S} = [X \mathsf{M}]$ consists of a set of tuples $(x,\mathsf{m})$, where $x \in X$ and $\mathsf{m} \in \mathsf{M}$. Let $\mathfrak{U}^\mathsf{M}_X$ be the corresponding set of compatible USTs:
$$ \mathfrak{U}^\mathsf{M}_X = \{\mathcal{U}^\mathsf{m}_x:\ (x,\mathsf{m}) \in \mathcal{S}\}.$$
Grammar $\mathcal{G}$ that concentrates probability mass on $\mathfrak{U}^\mathsf{M}_X$ can be estimated using the classical Bayesian approach:
$$
\mathcal{G}_* = \argmax_{\mathcal{G}} prob(\mathcal{G} \mid \mathfrak{U}^\mathsf{M}_X) = \argmax_{\mathcal{G}} \frac{prob(\mathcal{G}) \cdot prob(\mathfrak{U}^\mathsf{M}_X \mid \mathcal{G})}{prob(\mathfrak{U}^\mathsf{M}_X)}.
$$
Noting that $prob(\mathfrak{U}^\mathsf{M}_X)$ does not influence the result and, in the lack of prior knowledge, assuming $prob(\mathcal{G})$ uniformly distributed among all $\mathcal{G}$, the solution is then given by the maximum likelihood formula:  
$$\mathcal{G}_* = \argmax_{\mathcal{G}} prob(\mathcal{G}\mid\mathfrak{U}^\mathsf{M}_X) \simeq \mathcal{G}_\textrm{ML} = \argmax_{\mathcal{G}} prob(\mathfrak{U}^\mathsf{M}_X \mid \mathcal{G}).$$
Assuming independence of $\mathcal{U}^\mathsf{m}_x$s:
$$\mathcal{G}_\textrm{ML} = \argmax_{\mathcal{G}} \prod_{\mathcal{U}^\mathsf{m}_x \in \mathfrak{U}^\mathsf{M}_X} prob(\mathcal{U}^\mathsf{m}_x \mid \mathcal{G}) = \argmax_{\mathcal{G}} \prod_{(x,\mathsf{m})\in\mathcal{S}} \sum_{y \in \mathcal{Y}^\mathsf{m}_x} prob(y \mid \mathcal{G}).$$
In the absence of contact constraints the maximization problem becomes equivalent to the standard problem of estimating grammar $\mathcal{G}$ given the sample $X$:
$$\mathcal{G}^\mathsf{m=0}_{\textrm{ML}} = \argmax_{\mathcal{G}} \prod_{\mathcal{U}_x \in \mathfrak{U}_X} prob(\mathcal{U}_x \mid \mathcal{G}) =  \argmax_{\mathcal{G}} \prod_{x \in X} \sum_{y \in \mathcal{Y}_x} prob(y \mid \mathcal{G}),$$
where $\mathsf{m=0}$ denotes a square null matrix of size equal to the length of the corresponding sequence, and $\mathfrak{U}_X = \{\mathcal{U}^\mathsf{m=0}_x: x \in X\}$.

\paragraph{Contrastive estimation}
Often it is reasonable to expect that $\mathcal{U}_\mathrm{query}$ comes from a neighbourhood of the target population $\mathcal{N}(\mathfrak{U}_\mathrm{target}) \subset \mathcal{U}_*$. In such cases it is practical to perform \textit{contrastive estimation} \citep{Smith2005}, which aims at concentrating probability mass distributed by the grammar from the neighbourhood of the of sample $\mathcal{N}(\mathfrak{U}_\mathrm{sample})$ to the sample itself $\mathfrak{U}_\mathrm{sample}$, such that:
$$\mathcal{G}_{\mathrm{CE}} = \argmax_{\mathcal{G}} \prod_{\mathcal{U}_x \in \mathfrak{U}_\mathrm{sample}} \frac{ prob(\mathcal{U}_x \mid \mathcal{G})}{ prob(\mathcal{N}(\mathcal{U}_x) \mid \mathcal{G})}.$$
Consider two interesting neighbourhoods. First, assume that contact map $\mathsf{m}$ is known and conserved in the target population and hence in the sample: $\mathfrak{U}^\mathsf{m}_X = \{\mathcal{U}^\mathsf{m}_x: x\in X\}$. This implies the same length $n$ of all sequences. Then $\mathcal{U}^\mathsf{m}_n$ is a reasonable neighbourhood of the target population, so
$$\mathcal{G}_{\mathrm{CE}(\mathsf{m})} = \argmax_{\mathcal{G}} \prod_{\mathcal{U}^\mathsf{m}_x \in \mathfrak{U}^\mathsf{m}_X} \frac{ prob(\mathcal{U}^\mathsf{m}_x \mid \mathcal{G})}{ prob(\mathcal{U}^\mathsf{m}_n \mid \mathcal{G})} = \argmax_{\mathcal{G}} \frac{\prod_{x \in X} \sum_{y \in \mathcal{Y}^\mathsf{m}_x} prob(y \mid \mathcal{G})} {\left[\sum_{y\in\mathcal{Y}^\mathsf{m}_n} prob(y \mid \mathcal{G})\right]^{|X|}}.$$
Second, assume that sequence $x$ is known to be yielded by the target population and the goal is to maximize likelihood that shapes of parse trees generated with $\mathcal{G}$ are consistent with contact map $\mathsf{m}$. Then $\mathfrak{U}_X$ is a reasonable neighbourhood of the sample $\mathfrak{U}^\mathsf{M}_X$, so
$$\mathcal{G}_{\mathrm{CE}(X)} = \argmax_{\mathcal{G}} \prod_{(x,\mathsf{m})\in\mathcal{S}} \frac{ prob(\mathcal{U}^\mathsf{m}_x \mid \mathcal{G})}{ prob(\mathcal{U}_x \mid \mathcal{G})} = \argmax_{\mathcal{G}} \prod_{(x,\mathsf{m})\in\mathcal{S}} \frac{\sum_{y \in \mathcal{Y}^\mathsf{m}_x} prob(y \mid \mathcal{G})}{\sum_{y \in \mathcal{Y}_x} prob(y \mid \mathcal{G})}.$$

\subsection{Simple(r) instance}
\label{sec:simple-instance}

\subsubsection{Definitions}
\label{sec:simple-instance-definitions}

Let $\ddot{\mathcal{G}}=\langle \Sigma, V, v_0, R, \theta \rangle$ be a probabilistic context-free grammar such that $V = V_T \uplus V_N$, $R = R_a \uplus R_b \uplus R_c$, and
\begin{equation*}
\begin{array}{l}
R_a = \{r_i: V_T \rightarrow \Sigma \},\\
R_b = \{r_j: V_N \rightarrow (V_N \cup V_T)\ (V_N \cup V_T) \},\\
R_c = \{r_k: V_N \rightarrow V_T\ V_N\ V_T \}.
\end{array}
\end{equation*}
Subsets $R_a$, $R_b$ and $R_c$ are referred to as \textit{lexical}, \textit{branching}, and \textit{contact} rules, respectively. Joint subset $R_b \cup R_c$ is referred to as \textit{structural} rules. 

Let $\mathsf{m}$ be a contact matrix compatible with the context-free grammar, i.e. no pair of positions in contact overlaps nor crosses boundaries of other pairs in contact (though pairs can be nested one in another):
$$\forall(i,j)\ m_{i,j}=1 \land (i \leq k \leq j \oplus i \leq l \leq j) \Rightarrow m_{k,l}=0,$$
where $\oplus$ denotes the exclusive disjunction, and positions in contact are separated from each other by at least 2: 
$$\forall(i,j)\ i < j + 2.$$

Let distance threshold in tree $\delta=4$. Then a complete parse tree $y$ generated by $\ddot{\mathcal{G}}$ is consistent with $\mathsf{m}$ only if for all $m_{i,j}=1$ derivation
$$\alpha_{1,i-1}\ v_k\ \alpha_{j+1,n} \overset{*}{\Rightarrow} \alpha_{1,i-1}\ x_i\ v_l\ x_j\ \alpha_{j+1,n}$$
is performed with a string of production rules
$$[v_k\rightarrow v_t v_l v_u ][ v_t \rightarrow x_i ][ v_t \rightarrow x_j],$$
where $\alpha_{i,j} \in (\Sigma \cup V)^{j-i+1}$, $v_k, v_l \in V_N$ and $v_t, v_u \in V_T$.

\subsubsection{Parsing}
\label{sec:simple-instance-parsing}
Given an input sequence $x$ of length $n$ and a grammar $\ddot{\mathcal{G}}$, $prob(x \mid \ddot{\mathcal{G}}) \equiv prob(\mathcal{Y}_x \mid \ddot{\mathcal{G}}) = \sum_{y \in \mathcal{Y}_x} prob(y \mid \ddot{\mathcal{G}})$ can be calculated in $O(n^3)$ by a slightly modified probabilistic Cocke-Kasami-Younger bottom-up chart parser \citep{co69,ka65,yo67}. Indeed, productions in $R_a \uplus R_b$ conforms to the Chomsky Normal Form \citep{ch59}, while it is easy to see that productions in $R_c$ requires only $O(n^2)$. The algorithm computes $prob(x \mid \ddot{\mathcal{G}}) = prob(\mathcal{Y}_x \mid \ddot{\mathcal{G}})$ in chart table $\mathsf{P}$ of dimensions $n \times n\times |V|$, which effectively sums up probabilities of all possible parse trees $\mathcal{Y}_x$.
In the first step, probabilities of assigning lexical non-terminals $V_T$ for each terminal in the sequence $x$ are stored in the bottom matrix $\mathsf{P_1} = \mathsf{P}[1,:,:]$. Then, the table $\mathsf{P}$ is iteratively filled upwards with probabilities $ \mathsf{P}[j,i,v] = prob(v \overset{*}{\Rightarrow} x_i \dots x_{i+j-1} \mid v \in V, \ddot{\mathcal{G}})$. Finally, $prob(\mathcal{Y}^\mathsf{m}_x \mid \ddot{\mathcal{G}}) = \mathsf{P}[n,1,v_0]$.

New extended version of the algorithm (Fig.~\ref{fig:code}) computes $prob(\mathcal{Y}^\mathsf{m}_x \mid \ddot{\mathcal{G}})$, i.e. it considers only parse trees $\mathcal{Y}^\mathsf{m}_x$ which are consistent with $\mathsf{m}$. To this goal it uses an additional table $\mathsf{C}$ of dimensions $\sum(\mathsf{m})/2 \times n \times |V_T|$. After completing $\mathsf{P_1}$ (lines~10-12), probabilities of assigning lexical non-terminals $V_T$ at positions involved in contacts are moved from $\mathsf{P_1}$ to $\mathsf{C}$ (lines~13-21) such that each matrix $\mathsf{C}_p = \mathsf{C}[p,:,:]$ corresponds to $p$-th contact in $\mathsf{m}$. In the subsequent steps $\mathsf{C}$ can only be used to complete productions in $R_c$; moreover both lexical non-terminals have to come either from $\mathsf{P}_1$ or $\mathsf{C}$, they can never be mixed (lines~35-40). The computational complexity of the extended algorithm is still $O(n^3)$ as processing of productions in $R_c$ has to be multiplied by iterating over the number of contact pairs in $\mathsf{m}$, which is $O(n)$ since the cross-serial dependencies are not allowed.

\linespread{1.0}
\begin{figure}
\begin{verbatim}
01: function parse_cky_cm(x, m, Ra, Rb, Rc, Vt, Vn, v0)
02: # input:
03: # x - sequence, m - contact map
04: # Ra - lexical, Rb - branching, Rc - contact rules
05: # Vt - set of lexical, Vn - set of non-lexical non-terminals
06: # v0 - start symbol

07:     n = length(x)
08:     P[n, n, |Vn|+|Vt|] = 0.0
09:     C[sum(m)/2, n, |Vt|] = 0.0

10:     for i=1 to n
11:         for r in Ra
12:             if x[i]==r.rhs[1] P[1,i,r.lhs] = r.prob
13:     num_p=0
14:     for i=1 to n-2
15:         for j=i+2 to n
16:             if m[i,j]==1
17:                 for r in Ra
18:                     P[1,i,r.lhs] = P[1,j,r.lhs] = 0.0
19:                     if x[i]==r.rhs[1] C[p,i,r.lhs] = r.prob
20:                     if x[j]==r.rhs[1] C[p,j,r.lhs] = r.prob 
21:                 num_p=num_p+1
22:     for j=2 to n
23:         for i=1 to n-j+1
24:             for k=1 to j-1
25:                 for r in Rb
26:                     P[j,i,r.lhs] += r.prob
27:                                  * P[  k,i,  r.rhs[1]]
28:                                  * P[j-k,i+k,r.rhs[2]]
29:             if (j>=3)
30:                 for r in Rc
31:                     P[j,i,r.lhs] += r.prob
32:                                  * P[1,  i,  r.rhs[1]]
33:                                  * P[j-2,i+1,r.rhs[2]]
34:                                  * P[1,  i+j,r.rhs[3]]
35:                 for c=0 to num_p-1
36:                     for r in Rc
37:                         P[j,i,r.lhs] += r.prob
38:                                      * C[p,  i,  r.rhs[1]]
39:                                      * P[j-2,i+1,r.rhs[2]]
40:                                      * C[p,  i+j,r.rhs[3]]
41:     return P[n, 1, v0]
\end{verbatim}
\caption{Pseudocode of the modified CKY parser}
\label{fig:code}
\end{figure}
\linespread{1.5}

\subsubsection{Calculating $prob(\mathcal{U}^\mathsf{m}_n \mid \ddot{\mathcal{G}})$}
\label{sec:simple-instance-contrastive_to_m}
This section shows effective computing $prob(\mathcal{U}^\mathsf{m}_n \mid \ddot{\mathcal{G}})$, which is denominator for the contrastive estimation of $\mathcal{G}_{\mathrm{CE}(\mathsf{m})}$ (cf. section~\ref{sec:estimation}). Given a sequence $x$ of length $n$, a corresponding matrix $m$ of size $n \times n$ and a grammar $\ddot{\mathcal{G}}$ the probability of a set of trees consistent with $\mathsf{m}$ is
$$
prob(\mathcal{U}^\mathsf{m}_n \mid \ddot{\mathcal{G}})
\equiv \sum_{x\in\Sigma^{n}} prob(\mathcal{U}^\mathsf{m}_x \mid \ddot{\mathcal{G}})
= \sum_{x\in\Sigma^{n}} \sum_{y \in \mathcal{Y}^\mathsf{m}_x} prob(y \mid \ddot{\mathcal{G}}).
$$
Given grammar $\ddot{\mathcal{G}}$, any complete derivation $r$ is a composition $r = \dot{r} \circ \tilde{r}$, where $\dot{r} \in (R_a)^*$ and $\tilde{r} \in (R_b \cup R_c)^*$. Let $y$ be a parse tree corresponding to derivation $r$, and let $\tilde{y}$ be an incomplete parse tree corresponding to derivation $\tilde{r}$. Note that for any $y$ corresponding to $r = \dot{r} \circ \tilde{r}$ there exists one and only one $\tilde{y}$ corresponding to $\tilde{r}$. Let $\tilde{\mathcal{Y}}^\mathsf{m}_x$ denote the set of such incomplete trees $\tilde{y}$. Note that labels of the leaf nodes of $\tilde{y}$ are lexical non-terminals $\forall(i)\ \alpha_{i,i} \in V_T$, and that $\dot{r}$ represents the unique left-most derivation $yield(\tilde{y}) \overset{*}{\Rightarrow} x$. Thus,
$$
\sum_{x\in\Sigma^n} \sum_{y\in\mathcal{Y}^\mathsf{m}_x} prob(y \mid \ddot{\mathcal{G}})
= \sum_{x\in\Sigma^n} \sum_{\tilde{y}\in\tilde{\mathcal{Y}}^\mathsf{m}_x} prob(\tilde{y} \mid \ddot{\mathcal{G}}) \cdot prob(yield(\tilde{y})\overset{*}{\Rightarrow} x \mid \ddot{\mathcal{G}}).
$$
Note that value of the expression will not change if the second summation is over $\tilde{y}\in\tilde{\mathcal{Y}}^\mathsf{m}_n$ since $\forall(\tilde{y} \notin \tilde{\mathcal{Y}}^\mathsf{m}_x)\ prob(yield(\tilde{y})\overset{*}{\Rightarrow} x \mid \ddot{\mathcal{G}}) = 0$. Combining with observation that $prob(\tilde{y} \mid \ddot{\mathcal{G}})$ does not depend on $x$, the expression can be therefore rewritten as:
$$
\sum_{x\in\Sigma^n} \sum_{y\in\mathcal{Y}^\mathsf{m}_x} prob(y \mid \ddot{\mathcal{G}}) = \sum_{\tilde{y}\in\tilde{\mathcal{Y}}^\mathsf{m}_n} prob(\tilde{y} \mid \ddot{\mathcal{G}}) \cdot \sum_{x\in\Sigma^n} prob(yield(\tilde{y})\overset{*}{\Rightarrow} x \mid \ddot{\mathcal{G}}).
$$
However, if $\ddot{\mathcal{G}}$ is \textit{proper}, then $\forall(\tilde{y}\in\tilde{\mathcal{Y}}^\mathsf{m}_n)\ \sum_{x\in\Sigma^n} prob(yield(\tilde{y})\overset{*}{\Rightarrow} x \mid \ddot{\mathcal{G}}) = 1$, as:
\begin{equation*}
\begin{aligned}
&\sum_{x\in\Sigma^n} prob(yield(\tilde{y})\overset{*}{\Rightarrow} x \mid \ddot{\mathcal{G}}) = \sum_{x\in\Sigma^n} \prod_{i=1}^n \theta(\alpha_{i,i} \rightarrow x_i) =\\
&\sum_{x\in\Sigma^n} \theta(\alpha_{1,1} \rightarrow x_1) \cdot\ldots\cdot \theta(\alpha_{n,n} \rightarrow x_n) =\\
&\theta(\alpha_{1,1} \rightarrow a_1) \cdot \theta(\alpha_{2,2} \rightarrow a_1) \cdot\ldots\cdot  \theta(\alpha_{n-1,n-1} \rightarrow a_1) \cdot \theta(\alpha_{n,n} \rightarrow a_1)\ +\\
&\theta(\alpha_{1,1} \rightarrow a_1) \cdot \theta(\alpha_{2,2} \rightarrow a_1) \cdot\ldots\cdot  \theta(\alpha_{n-1,n-1} \rightarrow a_1) \cdot \theta(\alpha_{n,n} \rightarrow a_2)\ +\\
&\ \vdots\\
&\theta(\alpha_{1,1} \rightarrow a_{|\Sigma|}) \cdot \theta(\alpha_{2,2} \rightarrow a_{|\Sigma|})  \cdot\ldots\cdot \theta(\alpha_{n-1,n-1} \rightarrow a_{|\Sigma|}) \cdot \theta(\alpha_{n,n} \rightarrow a_{|\Sigma|}) =
\end{aligned}
\end{equation*}
\begin{equation*}
\left(
\begin{aligned}
&\theta(\alpha_{1,1} \rightarrow a_1) \cdot \theta(\alpha_{2,2} \rightarrow a_1) \cdot\ldots\cdot  \theta(\alpha_{n-1,n-1} \rightarrow a_1)\ + \\
&\theta(\alpha_{1,1} \rightarrow a_1) \cdot \theta(\alpha_{2,2} \rightarrow a_1) \cdot\ldots\cdot  \theta(\alpha_{n-1,n-1} \rightarrow a_2)\ + \\
&\ \vdots\\
&\theta(\alpha_{1,1} \rightarrow a_{|\Sigma|}) \cdot \theta(\alpha_{2,2} \rightarrow a_{|\Sigma|})  \cdot\ldots\cdot \theta(\alpha_{n-1,n-1} \rightarrow a_{|\Sigma|})
\end{aligned}
\right) \cdot \sum_{s=1}^{|\Sigma|} \theta(\alpha_{n,n} \rightarrow a_s),
\end{equation*}
where $a_s \in \Sigma$. Since $\ddot{\mathcal{G}}$ is \textit{proper} then $\forall(v \in V_T)\ \sum_{s=1}^{|\Sigma|} \theta(v \rightarrow a_s) = 1$ and therefore the entire formula evaluates to 1, which can be easily shown by iterative regrouping. This leads to the final formula:
$$
prob(\mathcal{U}^\mathsf{m}_n \mid \ddot{\mathcal{G}}) = \sum_{\tilde{y}\in\tilde{\mathcal{Y}}^\mathsf{m}_n} prob(\tilde{y} \mid \ddot{\mathcal{G}}).
$$
Technically, $\sum_{\tilde{y}\in\tilde{\mathcal{Y}}^\mathsf{m}_n} prob(\tilde{y} \mid \ddot{\mathcal{G}})$ can be readily calculated by the bottom-up chart parser by setting $\forall(r_k \in R_a)\ \theta(r_k)=1$.

\subsection{Evaluation}
\label{sec:evaluation}

\subsubsection{Learning}
\label{sec:evaluation-learning}
The present PCFG-CM approach was evaluated in practice for grammatical models $\ddot{G}$ and $\bar{G} = \ddot{G} \setminus R_c$ (the same grammar but without the contact rules) using an on-site framework for learning rule probabilities \citep{dy09, Dyrka2013}. Given an underlying CFG $\ddot{G}$, the framework estimates rule probabilities $\theta$ for the corresponding PCFG $\ddot{\mathcal{G}} = \langle \ddot{G}, \theta \rangle$ from the positive sample using a genetic algorithm in the Pittsburgh flavour, where each individual represents a whole grammar. Unlike previous applications of the framework in which probabilities of the lexical rules were fixed according to representative physicochemical properties of amino acids, in this research probabilities of all rules were subject to evolution. The objective functions were implemented for estimators $\ddot{\mathcal{G}}_\textrm{ML}$, $\ddot{\mathcal{G}}_{\mathrm{CE}(X)}$, and $\ddot{\mathcal{G}}_{\mathrm{CE}(\mathsf{m})}$. Besides, the setup of the genetic algorithm closely followed that of \citep{dy09}.

The input non-probabilistic grammar $\ddot{G}$ consisted of an alphabet of twenty terminal symbols representing amino acid species
$$\Sigma = \{A,C,D,E,F,G,H,I,K,L,M,N,Q,P,R,S,T,V,W,Y\},$$
a set of non-terminals symbols $V = V_T \uplus V_N$, where $V_T = \{l_1, l_2, l_3\}$ and $V_N = \{v_0, v_1, v_2, v_3\}$, and a set of rules $R = R_a \uplus R_b \uplus R_c$, which consisted of all possible allowed combinations of symbols, hence $|R_a| = 60, |R_b|=196, |R_c|= 144$. The set of non-contact rules was identical to the standard grammar in \citep{dy09}. The number of non-terminal symbols was limited to a few in order to keep reasonable the number of parameters to be optimized by the genetic algorithm. Combinations of symbols in rules were not constrained beyond general definition of the model $\ddot{G}$ in order to avoid interference with contact-map constraints, for the sake of transparent evaluation of the PCFG-CM.

\subsubsection{Performance measures}
\label{sec:evaluation-performance}
Performance of grammars was evaluated using a variant of the $k$-‐fold Cross‐-Validation scheme in which $k-2$ parts are used for training, $1$ part is used for validation and parameter selection, and $1$ part is used for the final testing and reporting results. Negative set was not used in the training phase. 

In order to avoid composition bias, proteins in the test sample were scored against the null model (encoded as a unigram), which assumed global average frequencies of amino acids, no contact information, and the sequence length of the protein. The amino acid frequencies were obtained using the online ProtScale tool for the UniProtKB/Swiss-Prot database \citep{Gasteiger2005}). 

\paragraph{Discriminative performance}
Grammars were assessed on the basis of the average precision (AP) in the recall-precision curve (RPC). The advantage of RPC over the more common Receiver Operating Characteristic (ROC) is robustness to unbalanced samples where negative data is much more numerous than positive data \citep{da06}. AP approximates the area under RPC.

\paragraph{Descriptive performance}
Intuitively, a decent explanatory grammar generates parse trees consistent with spatial structure of the protein. Perhaps the most straightforward approach to assess descriptive performance is to use the UST of the maximum likelihood parse tree as a predictor of spatial contacts between positions in sequence, parametrized by the cutoff $\delta$ on path length between the leaves. The natural threshold for grammar $\ddot{G}$, which is $\delta=4$ (the shortest distance between terminals generated by $R_b$ rules), was used for calculating the precision of contact prediction. In addition, AP of RPC, which sums up over all possible cutoffs, was computed to allow comparison with grammars without pairing rules. Eventually, the recall of the contact prediction at the threshold $\delta=4$ measured with regard to the partial contact map used in the training was used to assess the learning process.

\paragraph{Implementation}
The PCFG-CM parser and the Protein Grammar Evolution framework were implemented in C++ using GAlib \citep{Wall2005} and Eigen \citep{Guennebaud2010}. Performance measures were implemented in Python~2 \citep{vanrossum1991} using Biopython \citep{Cock2009}, igraph \citep{Csardi2006}, NumPy \citep{vanderWalt2011}, pyparsing \citep{McGuire2008}, scikit-learn \citep{Pedregosa2011} and SciPy \citep{Jones2001}.

\section{Results}
\label{sec:results}

\subsection{Materials}
\label{sec:results-materials}
Probabilistic grammars were estimated for three samples of protein fragments based on functionally relevant gapless motifs \citep{si02, bailey1994}. Within each sample, all sequences shared the same length, which avoided sequence length effects on grammar scores (which could be resolved by appropriate the null model). For each sample, one experimentally solved spatial structure in the Protein Data Bank (PDB) \citep{be00} was selected as a representative. Three samples included amino acid sequence of two small ligand binding sites (already analysed in \citep{dy09}) and one functional amyloid HET-s (Table~\ref{tab:datasets}):
\begin{itemize}
\item \textit{CaMn}: a Calcium and Manganese binding site from the legume lectins \citep{Sharon1990} collected according to the PROSITE PS00307 pattern \citep{Sigrist2013} true positive and false negative hits. Boundaries of the motif were extended to cover the entire binding site, similarly to \citep{dy09}. The motif folds into a stem-like structure with multiple contacts, many of them forming nested dependencies, which stabilize anti-parallel beta-sheet made of two ends of the motif (Fig.~\ref{fig:schematic}a based on pdb:2zbj \citep{deOliveira2008});
\item \textit{NAP}: the Nicotinamide Adenine dinucleotide Phosphate binding site fragment from an aldo/keto reductase family \citep{Bohren1989} collected according to the PS00063 pattern true positive and false negative hits (four least consistent sequences were excluded). The motif is only a part of the binding site of the relatively large ligand. The intra-motif contacts seems to be insufficient for defining the fold, which depends also on interactions with amino acids outside the motif (Fig.~\ref{fig:schematic}b based on pdb:1mrq \citep{Couture2003});
\item \textit{HET-s}: the HET-s-related motifs r1 and r2 involved in the prion-like signal transduction in fungi identified in a recent study \citep{Daskalov2015}. The largest subset of motifs with length of 21 amino acids was used to avoid length effects on grammar scores. When interacting with a related motif r0 from a cooperating protein, motifs r1 and r2 adopt beta-hairpin-like folds which stack together. While stacking of multiple motifs from several proteins is essential for stability of the structure, interactions between hydrophobic amino acids within single hairpin are also important. In addition, correlation analysis revealed strong dependency between positions 17 and 21 \citep{Daskalov2015} (Fig.~\ref{fig:schematic}c based on \citep{vanMelckebeke2010}).
\end{itemize}
Diversity of sequences ranged from the most homogenous CaMn to the most diverse HET-s, which consisted of 5 subfamilies \citep{Daskalov2015}.

\begin{table}
\label{tab:datasets}
\caption{Datasets. Notations: \textit{sim} - maximum sequence similarity, \textit{npos/nneg} - number of positive/negative sequences, \textit{len} - sequence length in amino acids, \textit{ncon} - total number of non-local contacts (separation 3+), \textit{msiz} - number of contacts selected for training}
\begin{center}
\begin{tabular}{llrrrrlrr}
\hline
\multicolumn{1}{l}{id}&\multicolumn{1}{l}{type}&\multicolumn{1}{r}{sim}&\multicolumn{1}{r}{npos}&\multicolumn{1}{r}{nneg}&\multicolumn{1}{r}{len}&\multicolumn{1}{l}{pdb}&\multicolumn{1}{r}{ncon}&\multicolumn{1}{r}{msiz}\\[2pt]
\hline
CaMn    & binding-site  & 71\%  & 24    & 28560 & 27  & 2zbj    & 41 & 6\\
NAP     & binding-site  & 70\%  & 64    & 47736 & 16  & 1mrq    & 11 & 2\\
HET-s   & amyloid       & 70\%  & 160   & 33248 & 21  & 2kj3    & 10 & 3\\[2pt]
\hline
\end{tabular}
\end{center}
\end{table}

Negative samples were designed to roughly approximate the entire space of protein sequences. They were based on the  negative set from \citep{dy09}, which consisted of 829 single chain sequences of 300-500 residues retrieved from the Protein Data Bank \citep{be00} at identity of 30\% (accessed on 12th December 2006). For each positive sample, the corresponding negative sample was obtained by cutting the basic negative set into subsequences of the length of the positive sequences.

All samples were made non-redundant at level of sequence similarity around 70\%. Contact pairings were assigned manually and collectively to all sequences in the set based on a selected available spatial structure of a representative positive sequence in the PDB database (Fig.~\ref{fig:schematic}). 

\linespread{1.0}
\begin{figure}
\begin{verbatim}
a) CaMn             b) NAP                   c) HET-s
           
  DI                 NFNR                        G
 T--G               S \QR                       K E
 N--D               V--LE                      G--S
P    PN             G                         V   R
Y    Y              I                         V--V
 S  P               S                        T    L
 D--H               K                        E   I \E
 L  Y               A                         V   GN
 E--G               L                          SN
 V  I                                          T
 A--D               >1mrq:A159-174             T
 V  I               LAKSIGVSNFNRRQLE        
 I--K                                       >2kj3:A260-280
                                            TTNSVETVVGKGESRVLIGNE
>2zbj:A4-30
IVAVELDSYPNTDIGDPNYPHIGIDIK
\end{verbatim}
\caption{Schematic representation of structure of the sample motifs. Context-free-compatible contact pairings selected in this study are marked with dashes and slashes. Order of amino acids in sequence and its coordinates in protein are given below the structure. Notes: 1) in CaMn, only 4 out of 7 real hydrogen bond-related contacts in the stem-like part were included in the contact map for the sample for the sake of simplicity; 2) in HET-s, e.g. a pair V5 and I18 conforms to definition of contact, however it crosses another contact between L17 and E21.}
\label{fig:schematic}
\end{figure}
\linespread{1.5}

\subsection{Performance}
\label{sec:results-performance}

Probabilistic grammars with the contact rules $\ddot{\mathcal{G}}$ were learned through estimation of probabilities of rules $\theta$ for non-probabilistic CFG $\ddot{G}$ using input samples made of sequences coupled with the contact map $\mathfrak{U}_X^\mathsf{m}$, or using sequences alone $\mathfrak{U}_X$. Probabilistic grammars without the contact rules $\bar{\mathcal{G}}$ were learned using only the input samples made of sequences $\mathfrak{U}_X$, since they cannot generate parse trees consistent with contact maps at the distance threshold $\delta=4$. Note that since there is the one-to-one correspondence between input sample set $S=[X\mathsf{M}]$ and sample of UST sets $\mathfrak{U}^\mathsf{M}_X$, notations developed for the sets of USTs are used to denote the input samples.

\subsubsection{Discriminative power}
\label{sec:results-performance-discriminative}
For evaluation of discriminative power of the PCFG-CM approach, rule probabilities were estimated using the maximum-likelihood estimator (denoted ML) and the contrastive estimator with regard to the contact map (denoted CE(m)). Discriminative performance of the resulting probabilistic grammars on $\mathfrak{U}_X$ and $\mathfrak{U}_X^\mathsf{m}$ is presented in Tab.~\ref{tab:discriminative} in terms of the average precision. 

\begin{table}
\caption{Discriminative performance of grammars in terms of AP}
\label{tab:discriminative}
\begin{center}
\begin{tabular}{lrrrrrrr}
\hline
\multicolumn{1}{l}{Grammar}&\multicolumn{1}{l}{$\bar{\mathcal{G}}_\textrm{ML}^\mathsf{m=0}$}&\multicolumn{2}{l}{$\ddot{\mathcal{G}}_\textrm{ML}^\mathsf{m=0}$}&\multicolumn{2}{l}{$\ddot{\mathcal{G}}_\textrm{ML}$}&\multicolumn{2}{l}{$\ddot{\mathcal{G}}_{\textrm{CE}(\mathsf{m})}$}\\
\multicolumn{1}{l}{Test sample}&\multicolumn{1}{l}{$\mathfrak{U}_X$}&\multicolumn{1}{l}{$\mathfrak{U}_X$}&\multicolumn{1}{l}{$\mathfrak{U}^\mathsf{m}_X$}&\multicolumn{1}{l}{$\mathfrak{U}_X$}&\multicolumn{1}{l}{$\mathfrak{U}^\mathsf{m}_X$}&\multicolumn{1}{l}{$\mathfrak{U}_X$}&\multicolumn{1}{l}{$\mathfrak{U}^\mathsf{m}_X$}\\[2pt]
\hline
CaMn   & 0.94 & 0.96 & 0.67 & 0.95 & 0.95 & 0.79 & 0.98\\
NAP    & 0.78 & 0.86 & 0.28 & 0.75 & 0.79 & 0.24 & 0.91\\
HET-s  & 0.46 & 0.43 & 0.24 & 0.60 & 0.81 & 0.23 & 0.94\\
\hline
\end{tabular}
\end{center}
\end{table}

The baseline here is the average precision of grammars estimated without contact constraints, $\bar{\mathcal{G}}_\textrm{ML}^\mathsf{m=0}$ and $\ddot{\mathcal{G}}_\textrm{ML}^\mathsf{m=0}$, tested on sequences alone $\mathfrak{U}_X$, which ranged from 0.43-0.46 for HET-s to 0.94-0.96 for CaMn. The scores show negative correlation with diversity of the samples and limited effect of adding contact rules (though the latter may result from worse learning of increased number of parameters with added rules).
$\ddot{\mathcal{G}}_\textrm{ML}^\mathsf{m=0}$ performed much worse when tested on the samples with the contact map $\mathfrak{U}_X^\mathsf{m}$, which indicates that preference for parses consistent with $\mathsf{m}$ is at best limited when training without constraints.

For all three samples, the highest AP (0.91-0.98) achieved grammars obtained using the contrastive estimation with regard to the contact map $\ddot{\mathcal{G}}_{\textrm{CE}(\mathsf{m})}$ tested on the samples with the map $\mathfrak{U}_X^\mathsf{m}$. The improvement relative to the baseline was most pronounced for HET-s, yet still statistically significant ($p<0.05$) for NAP. As expected, the contrastively estimated grammars performed poorly on sequences alone $\mathfrak{U}_X$ except for the CaMn sample.

The maximum-likelihood grammars estimated with the contact information $\ddot{\mathcal{G}}_\textrm{ML}$ tested on $\mathfrak{U}_X^\mathsf{m}$ performed worse than the contrastively estimated grammars but comparably or significantly better (HET-s) than the baseline. The average precision of $\ddot{\mathcal{G}}_\textrm{ML}$ was consistently lower when tested without the map on sequences alone, yet still considerable (from 0.60 for HET-s to 0.95 for CaMn). It is notable that in the HET-s case, $\ddot{\mathcal{G}}_\textrm{ML}$ achieved better AP on $\mathfrak{U}_X$ than $\ddot{\mathcal{G}}_\textrm{ML}^\mathsf{m=0}$.

Notably high AP for CaMn with $\ddot{\mathcal{G}}_\textrm{ML}^\mathsf{m=0}$ tested on $\mathfrak{U}_X^\mathsf{m}$ and with $\ddot{\mathcal{G}}_{\textrm{CE}(\mathsf{m})}$ tested on $\mathfrak{U}_X$ can be contributed to relatively strong signal from the long stem-like part of the motif particularly suitable for modeling with the contact rules. 

\subsubsection{Descriptive power}
\label{sec:results-performance-descriptive}
For evaluation of descriptive power of the PCFG-CM approach, rule probabilities were estimated using the maximum-likelihood estimator (denoted ML) and the contrastive estimator with regard to sequences (denoted CE(X)). Descriptive value of the most probable parse trees generated using the resulting probabilistic grammars for test sequences without contact information $\mathfrak{U}_X$ is presented in Tab.~\ref{tab:explanatory}. Efficiency of the learning was measured on the basis of the recall at $\delta=4$ with regard to the context-free compatible contact map used in the training. Consistency of the most likely parse tree with the protein structure was measured on the basis of the precision of the contact prediction at $\delta=4$ with regard to all contacts in the reference spatial structure with separation in sequence of at least 3. Both measures are not suitable for assessing grammars without contact rules $\bar{\mathcal{G}}$. Therefore, average precision over all thresholds $\delta$ was used as secondary measure for consistency of the most likely trees with the protein structure. Note that the AP scores achievable for a context-free parse tree are reduced by overlapping pairings. 

\begin{table}
\caption{Descriptive quality of the most likely parse trees derived from sequences only, in terms of recall for $\delta=4$ w.r.t the known contact map $\mathsf{m}$, and precision for $\delta=4$ (and AP over thresholds $\delta$) w.r.t the full contact map of the reference \textit{pdb} structure for sequence separation 3+. Note that lengths of the shortest paths between leaves in the most likely parse trees of grammars $\bar{G}$ equal 5, which makes measures based on $\delta=4$ unutile.}
\label{tab:explanatory}
\begin{center}
\begin{tabular}{lrrrrrrr}
\hline
\multicolumn{1}{l}{Gram.}&\multicolumn{1}{l}{$\bar{\mathcal{G}}_\textrm{ML}^\mathsf{m=0}$}&\multicolumn{2}{l}{$\ddot{\mathcal{G}}_\textrm{ML}^\mathsf{m=0}$}&\multicolumn{2}{l}{$\ddot{\mathcal{G}}_\textrm{ML}$}&\multicolumn{2}{l}{$\ddot{\mathcal{G}}_{\textrm{CE}(X)}$}\\
\multicolumn{1}{l}{Ref.} & \multicolumn{1}{l}{pdb} & \multicolumn{1}{l}{$\mathsf{m}$} & \multicolumn{1}{l}{pdb} & \multicolumn{1}{l}{$\mathsf{m}$} & \multicolumn{1}{l}{pdb} & \multicolumn{1}{l}{$\mathsf{m}$} & \multicolumn{1}{l}{pdb}\\[2pt]
\hline
CaMn    & (0.24) & 0.45 & 0.69 (0.53) & 0.92 & 0.87 (0.66) & 0.98 & 0.84 (0.66)\\
NAP     & (0.16) & 0.00 & 0.14 (0.12) & 0.96 & 0.64 (0.29) & 0.96 & 0.64 (0.29)\\
HET-s   & (0.08) & 0.02 & 0.13 (0.14) & 0.79 & 0.52 (0.24) & 0.97 & 0.57 (0.27)\\
\hline
\end{tabular}
\end{center}
\end{table}

The baseline here are the results for grammars with the contact rules estimated without contact constraints $\ddot{\mathcal{G}}_\textrm{ML}^\mathsf{m=0}$. The most likely parse trees generated using these grammars conveyed practically no information about contacts for NAP and HET-s (recall w.r.t $\mathsf{m}$ close to zero) and limited information about contacts for CaMn (moderate recall of 0.45). Increase of the recall to 0.79-0.98 obtained for the most likely parse trees generated using grammars $\ddot{\mathcal{G}}_\textrm{ML}$ and $\ddot{\mathcal{G}}_{\textrm{CE}(X)}$ testifies efficiency of the learning process with contact constraints.

Importantly, consistency of the most likely parse trees with the protein structure measured by the precision followed a similar pattern and increased from 0.13 for HET-s, 0.14 for NAP, and 0.69 for CaMn when learning on $\mathfrak{U}_X$, to respectively 0.52-0.57, 0.64, and 0.84-0.87, when learning on $\mathfrak{U}_X^\mathsf{m}$. Accordingly, evaluation in terms of the average precision over distance thresholds indicated that distances in the most likely parse trees better reflected the protein structure if grammars were trained with the contact constraints on $\mathfrak{U}_X^\mathsf{m}$.

\section{Discussion and conclusions}
\label{sec:discussion}

\subsection{Analysis of computational results}
Computational validation of discriminatory power showed that additional knowledge present in the partial contact map can be effectively incorporated into the probabilistic grammatical framework through the concept of syntactic tree consistent with the contact map. The most effective way of training descriptors for a given sample was the contrastive estimation with reference to the contact map. This approach is only possible when a single contact map that fits all sequences in the target population can be used with the trained grammar. The maximum-likelihood estimators were effective when contacts were relevant to structure of the sequence (HET-s, CaMn). This is expected, as use of the contact rules is likely to be optimal for deriving a pair of amino acids in contact if they are actually correlated. Interestingly, in the case of HET-s, the maximum-likelihood grammar $\ddot{\mathcal{G}}_\textrm{ML}$ trained with the contact constraints compared favourably with the maximum-likelihood grammar $\ddot{\mathcal{G}}^\mathsf{m=0}_\textrm{ML}$ trained on sequences alone even when tested on sequences alone (AP 0.60 versus 0.43). In other words, $\ddot{\mathcal{G}}_\textrm{ML}$ was more optimal with regard to the sample of sequences than $\ddot{\mathcal{G}}^\mathsf{m=0}_\textrm{ML}$. This indicates that if contacts are relevant for the structure of sequence, the PCFG-CM approach can improve robustness of learning to local optima.

Computational validation of descriptive power showed that the most likely parse trees, derived for inputs defined only by sequences, reproduced vast majority of contacts (recall of at least 0.79 at $\delta=4$) enforced by the contact-map constrained training input. Moreover, precision of contact prediction at $\delta=4$ and sequence separation 3+ was above 0.50, up to 0.87. This translated to the overall overlap with the full contact maps in range 0.27-0.39 (not shown), since only a fraction of contacts can be represented in the parse tree of the context-free grammar, and even not all of them were enforced in the training or were optimally parsed with the contact rules. The benefit of contrastive estimation with reference to sequences was limited in comparison to maximum likelihood grammars. However, it should be noted that the shape of the most likely parse tree, which was used in the evaluation, does not necessarily reflects the most likely shape of parse tree. Unfortunately, the latter cannot be efficiently computed \citep{Dowell2004}. 

Reasonable performance of grammars estimated with contact constraints $\ddot{\mathcal{G}}_\textrm{ML}$ on sequences alone (AP from 0.60 to 0.95) is encouraging as it gives a hint of performance of the PCFG-CM approach in its potential most general application to model very diverse data sets where each training sequence is associated with a different contact map. In this case, contact maps cannot be used for recognizing unknown sequences. So far conclusive results for this kind of application are not yet available.

\subsection{Limitations and perspectives}
The computational experiments mainly served assessing intuitions, which led to development of the PCFG-CM approach. Full scale practical application to bioinformatic problems such as sequence search would certainly require several enhancements. For example, accurate accounting for various sequence length would likely require a more elaborated null model. Moreover, to increase the number of non-terminal symbols, the learning framework have to be improved. This includes more efficient estimation of probabilities of a large number of rules and/or added capability of inferring rules during learning \citep{Unold2005,Unold2012,cos12,cos14}. In addition, preliminary testing (not shown) suggests that scoring inputs with the product of probabilities using grammars with lexical rule probabilities fixed according to representative physicochemical properties of amino acids \citep{dy09}, and the appropriately adjusted null model, has more discriminative power than the current approach. Extension of the PCFG-CM framework to account for uncertain contact information as in \citep{Knudsen2005} seems to be straightforward through introducing the concept of fuzzy sets of syntactic trees. These application-related developments are left for future work.

Though tested in the learning setting consisting in optimizing only rule probabilities, the estimators defined in the present PCFG-CM framework can be used in more general learning schemes inferring also the grammar structure. Indeed, such schemes may even more benefit from constraining the larger search space. It is also interesting to consider extending the framework beyond context-free grammars as contacts in proteins are often overlapping and thus context-sensitive. In this case, however, the one-to-one correspondence between the parse tree and derivation breaks, therefore it may be advisable to redefine the grammatical counterpart of the spatial distance in terms of derivation steps in order to take advantage from higher expressiveness.

\paragraph{Contributions}
The PCFG-CM framework was proposed by WD and elaborated by WD, FC and JT. Implementation of the special instance, computational experiments and analysis of results were carried out by WD. The paper was written by WD, FC and JT.

\paragraph{Acknowledgements}
This research has been partially funded by National Science Centre, Poland [grant no 2015/17/D/ST6/04054] and was supported by the E-SCIENCE.PL Infrastructure. Computational experiments have been partially carried out using resources provided by Wroclaw Centre for Networking and Supercomputing (http://wcss.pl) [grant no 98]. Neither of the funding bodies influenced in any way the study design and collection, analysis and interpretation of data. Nor did they participate in writing of the manuscript. 

WD acknowledges Olgierd Unold and Mateusz Pyzik for interesting discussions in the course of the project.
%
%

\bibliographystyle{abbrvnat}
\bibliography{references}

\clearpage
\end{document}